\documentclass[10pt]{IEEEtran}

\usepackage{float}
\usepackage{subfigure}
\usepackage{stfloats}
\usepackage{cite}
\usepackage{amsmath,amssymb,amsfonts}
\usepackage{graphicx}
\usepackage{textcomp}
\usepackage{mathrsfs}
\usepackage{cases} 
\usepackage{algorithm}
\usepackage{algorithmicx}
\usepackage{algpseudocode}
\usepackage{color}
\usepackage{amsmath}

\newcommand{\be}{\begin{equation}}
\newcommand{\ee}{\end{equation}}
\def\bee#1\eee{\begin{align}#1\end{align}}
\newcommand{\bse}{\begin{subequations}}
\newcommand{\ese}{\end{subequations}}

\ifCLASSINFOpdf
\else
\fi

\begin{document}
	
	\title{Technical Report for A Service-Fair UAV-NOMA System for Large-scale IoT scenarios}

	\author{Xindi Wang}

\maketitle

\IEEEpeerreviewmaketitle

\section{The smallest enclosing circle Algorithm}
Suppose there are $k$ points $P=\{p_i=[x_i,y_i]\}_{i=1}^k$ in the scene, and we need to find a circle with the smallest radius to cover these points, regarded as the smallest enclosing circle for these $k$ points.

\textbf{Proposition 1}: The smallest enclosing circle for $k$ points should pass through at least two points.

\textbf{Proof:} It is straightforward to prove it.

\textbf{Proposition 2}: The smallest enclosing circle for $k$ points is generated by at most three points.

\textbf{Proof:} The proof is given in \cite{Welzl}.

\textbf{Proposition 3}: If point $k$ is outside of the smallest enclosing circle for points $\{p_1,p_2,\dots,p_{k-1}\}$, then the point $k$ should be on the boundary of the smallest enclosing circle for points $\{p_1,p_2,\dots,p_k\}$.

\textbf{Proof:} The proof is given in \cite{Welzl}.

Then, we divide the searching problem of the smallest enclosing circle into two cases to discuss separately.

$\blacksquare$ Case $1$: The smallest enclosing circle for $k$ points pass through only two points of $P$.

If the circle passes through two points (denoted as $p_i$ and $p_j$) of $P$, then the line connecting these two points is the diameter of the circle. In this case, the radius and center can be given by $\textbf{L}_c=\frac{p_i+p_j}{2}$ and $r_c=\frac{||p_i-p_j||}{2}$.

$\blacksquare$ Case $2$: The smallest enclosing circle for $k$ points should pass through at least three points of $P$.

If the smallest enclosing circle passes through three points (denoted as $p_i$, $p_j$ and $p_h$) of $P$, then the center of the circle (i.e. $\textbf{L}_c=[x_0,y_0]$) can be given by
\begin{numcases}{}
	x_0=\frac{de-bf}{bc-ad},\\
	y_0=\frac{af-ce}{bc-ad}.
\end{numcases}
where
\begin{numcases}{}
	a=x_i-x_j,\nonumber\\
	b=y_i-y_j,\nonumber\\
	c=x_i-x_h,\nonumber\\
	d=y_i-y_h,\nonumber\\
	e=\frac{(x_i^2-x_j^2)-(y_j^2-y_i^2)}{2},\nonumber\\
	f=\frac{(x_i^2-x_h^2)-(y_h^2-y_i^2)}{2}.\nonumber
\end{numcases}
, and the radius is given by $\textbf{L}_c=\frac{[x_0,y_0]+[x_i,y_i]}{2}$.

Assuming that the points in $P$ are with weights $\{W_i|W_1\geq W_2\geq \dots\geq W_k\}_{i=1}^k$, respectively. If we need to find the smallest enclosing circle for at most $n$ points of $P$, which owns the maximum sum of the weight value, an efficient and effective method is given as follows.

\begin{algorithm}
	\caption{Greedy Search for the Constrained Smallest Enclosing Circle}
	\label{alg:sp1}
	\begin{algorithmic}[1] 
		\State $\Omega_i^*=\{p_1, p_2\}$, $\textbf{L}_c=\frac{p_1+p_2}{2}$, $r_c=\frac{||p_1-p_2||}{2}$, k=3.
		\While{$|\Omega_i^*|\leq n$}
		\State \textbf{If} $||p_k-\textbf{L}_c||\leq r_c$:
		\State \ \ \ $\Omega_i^*=\Omega_i^*\cup p_k$.
		\State \ \ \ Search the remaining points from $P/ \Omega_i^*$, and add them into $\Omega_i^*$. 
		\State \textbf{Else}:
		\State \ \ \ \ \ Construct a circle $c$ whose diameter is the line connecting from point $p_k$ to the point that is the furthest from $p_k$ from the set $P/ \Omega_i^*$. 
		\State \ \ \ \ \ \textbf{If} all points of $\Omega_i^*$ is within the circle $c$:
		\State \ \ \ \ \ \ \ Update $\textbf{L}_c$ and $r_c$.
		\State \ \ \ \ \ \textbf{Else}:
		\State \ \ \ \ \ \ \ \ \textbf{For} each two points in $\Omega_i^*$:
		\State \ \ \ \ \ \ \ \ \ \ \ Construct a circle based on the two points and $p_k$.
		\State \ \ \ \ \ \ \ \ \ \ \ \ \ Pick out the circle with the smallest radius, and update $\textbf{L}_c$ and $r_c$.
		\State \ \ \ \ \ \ \ \ \ \ \ \ \ Search the remaining points from $P/ \Omega_i^*$, and add them into $\Omega_i^*$.
		\EndWhile
		\State Retain the $n$ points with the largest weight value in $\Omega_i^*$, and output $\textbf{L}_c$ and $r_c$.
	\end{algorithmic}
\end{algorithm}

Due to the fact that the Algorithm.~1 is to search the smallest enclosing circle for partial points in a greedy way, thus the results of the Algorithm.~1 can be regard as a periodic achievement of the algorithm proposed in \cite{Welzl}. Therefore, as proven in \cite{Welzl}, the expectation of the computational complexity of the algorithm is given by $O(k^3)$, in which $k$ denotes the number of points.

\section{Extension in Probabilistic LoS Channels}
Here we discuss how to further extend this letter to probabilistic LoS channel scenarios. In this case, the waypoint of the UAV in time slot $t$ is denoted as   $\textbf{L}_u(t)=[x(t),y(t),z]^{T}\in\mathbb{R}^{3\times 1}$ in a given 3-D space. Then, the LoS probability at each time slot $t$ can be modeled as a function of the UAV-SN elevation angle, which can be expressed in the form of 
\begin{align}
	P(c^{L}_{k,t}=1)=\frac{1}{1+\alpha e^{-b(\theta_{k,t}-\alpha)}},
\end{align}
where $\theta_{k,t}$ denotes the angel between the UAV and SN $k$ in time slot $t$, which is given by
\begin{align}
	\theta_{k,t}=\frac{180}{\pi}arctan(\frac{z}{\sqrt{(x_i-x(t))^2+(y_i-y(t))^2}}),
\end{align}
where $\alpha$ and $\beta$ are modeling parameters to be specified. Then, the corresponding NLoS probability can be obtained by 
\begin{align}
	P(c^{L}_{k,t}=0)=1-P(c^{L}_{k,t}=1).
\end{align}

In this case, the large-scale channel power gain between the UAV and SN $k$ in the time slot $t$, including both the path loss and shadowing, can be modeled as 
\begin{align}\label{eq:h}
	h_{k,t}=c^{L}_{k,t}h^{L}_{k,t}+(1-c^{L}_{k,t}) h^{N}_{k,t},
\end{align}
where 
\begin{align}
	h^{L}_{k,t}=\beta_0 d^{-\alpha_L}_{k,t},\ h^{N}_{k,t}=\mu\beta_0 d^{-\alpha_N}_{k,t}
\end{align}
denote the channel power gains in LoS and NLoS cases, respectively. Meanwhile, $d_{k,t}$ can be calculated as
\begin{align}
	d_{k,t}=||\textbf{L}_i-\textbf{L}_u(t)||.
\end{align}

As a result, it only needs to replace $g_i(t)$ (i.e. (1) of the manuscript) with \eqref{eq:h} of this technical report, and the optimization problem (i.e. problem P0 of the manuscript) stays the same. The UAV-NOMA system with probability LoS channels can still be solved by the proposed schemes in the manuscript.

\bibliographystyle{IEEEtran}

\begin{thebibliography}{1}
\bibitem{Welzl}
E. Welzl, ``Smallest enclosing disks (balls and ellipsoids)''. \emph{New Results and New Trends in Computer Science. Lecture Notes in Computer Science}, vol 555. Springer, Berlin, Heidelberg, 1991.	
	
	

\end{thebibliography}

\end{document}